\documentclass[a4paper,11pt]{article}
\usepackage{pos}

\newcommand{\bmin}{\begin{minipage}{0.5\textwidth}}
\newcommand{\emin}{\end{minipage}}
\newcommand{\bmini}[1]{\begin{minipage}{#1}}
\newcommand{\bc}{\begin{center}}
\newcommand{\ec}{\end{center}}
\newcommand{\HUp}{\bf H}
\newcommand{\SU}{S({\bf U})}
\newcommand{\GU}{G({\bf U}_\mu)}
\def\bp{{\bf p_\mu}}
\newcommand{\MU}{M({\bf U})}

\title{Riemannian Manifold HMC with fermions}

\author*[a]{Chulwoo Jung}
\author[b]{Norman H. Christ}

\affiliation[a]{Brookhaven National Laboratory,\\
  New York, U.S.A. }

\affiliation[b]{Columbia University, New York, New York, U.S.A}

\emailAdd{chulwoo@bnl.gov}

\abstract{
We report on our study of the Riemannian Manifold HMC (RMHMC) algorithm with the mass term for the gauge momenta replaced by rational functions of the gauge covariant Laplace operator.
A comparison of HMC and RMHMC on a 2+1+1 flavor dynamical ensemble with lattice spacing $a \sim $0.05fm
shows increased rate of change in long distance modes, identified by Wilson flowed energy, 
per fermion molecular dynamics step.
}

\FullConference{The 40th International Symposium on Lattice Field Theory (Lattice 2023)\\
July 31st - August 4th, 2023\\
Fermi National Accelerator Laboratory\\}

\begin{document}
\maketitle

\section{Introduction}

A major source of the systematic error with LQCD is finite lattice spacing ($a$),
which can be controlled by simulating at successively smaller lattice spacings. However,
the increasing separation in size between the physical scale and
the lattice spacing means smaller integration time steps are needed to integrate short distance modes. 
One of the consequences of this is a rapidly increasing autocorrelation time in molecular dynamics(MD) units for ensemble generation at smaller lattices spacings.
For example, it was observed that the integrated
autocorrelation time for the topological charge $Q(\tau)$ scales as $a^{-5}$ or worse\cite{Sommer}, which means, when other expected scaling costs are included,
the numerical cost to generate the same number of decorrelated configurations increases as $\sim a^{-10}$ or more for the same physical volume.

\begin{figure}
\bc
\bmini{0.55\textwidth}
\includegraphics[width=\textwidth]{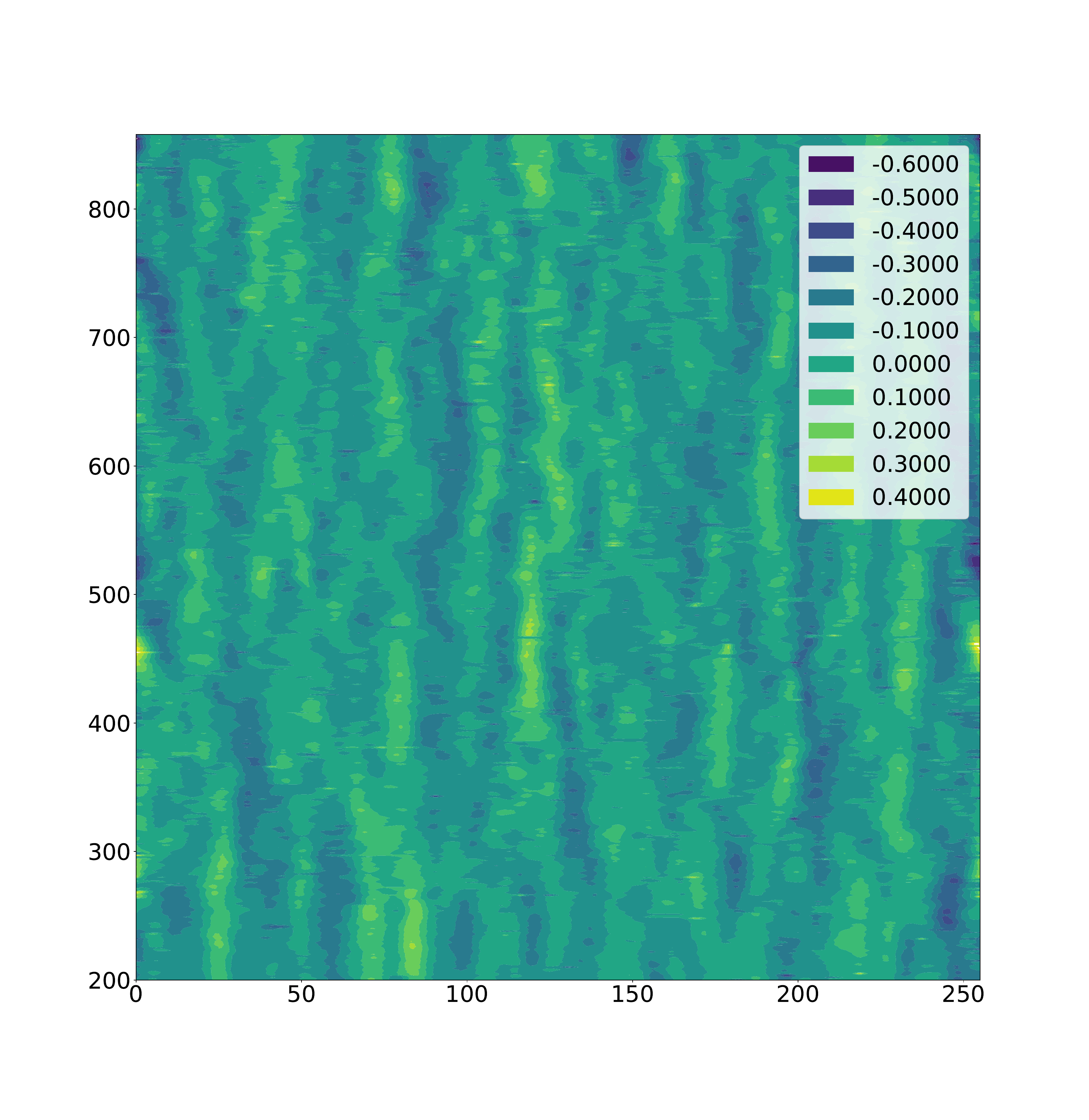}
\emin
\bmini{0.45\textwidth}
\includegraphics[width=\textwidth]{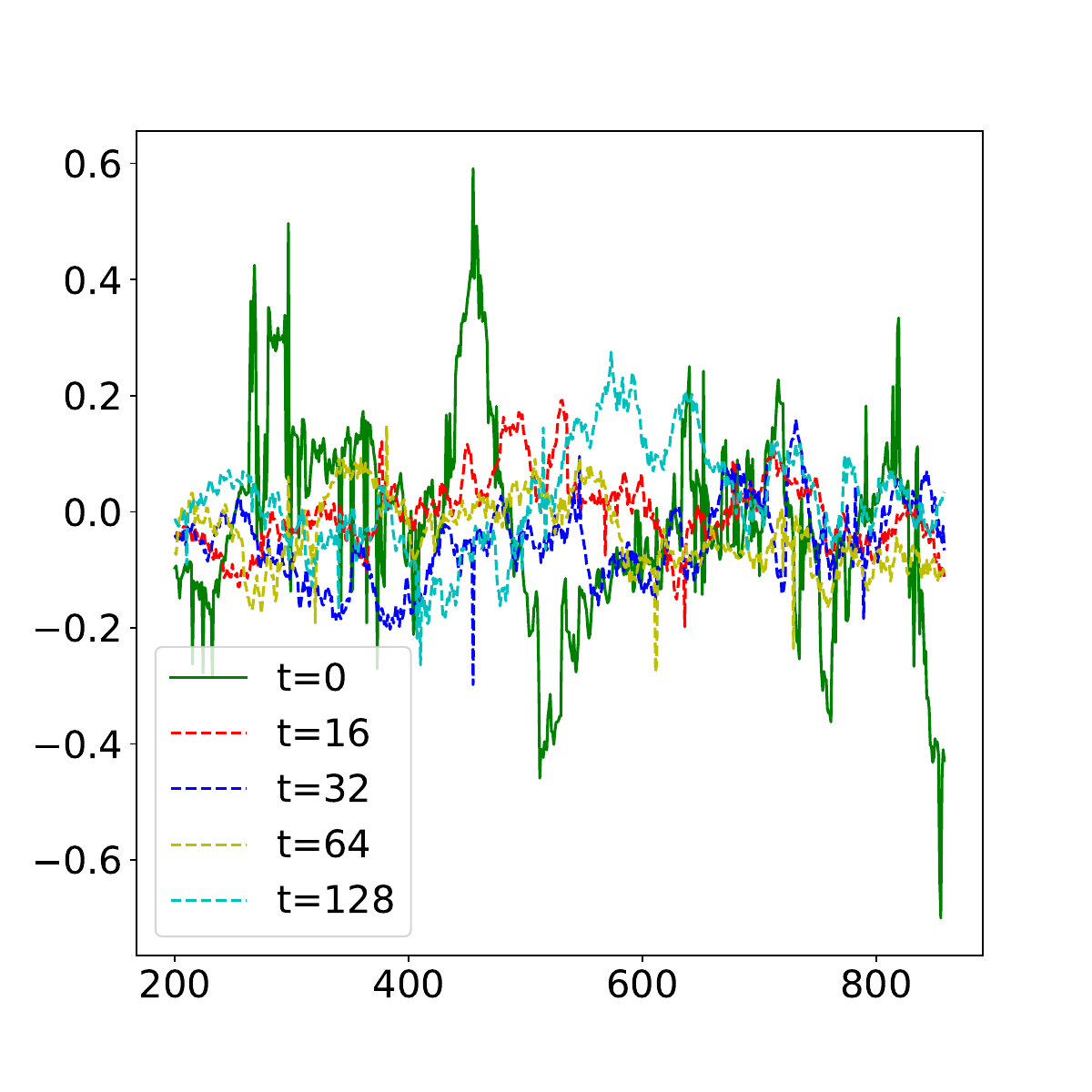}
\emin
\ec
	\caption{ Contour plot and MD time evolution of the topological charge per timeslice($Q(t,\tau)$)  for $32^3 \times 256, 1/a\sim$4Gev, 2+1+1-flavor DWF ensemble generated with open boundary condition in time directions. In the contour plot(left), the x axis is the time coordinate and the y axis is the trajectory number. In the MD time evolution plot(right), each line represents a different timeslice.}
\label{fig:top_t}
\end{figure}

Recently, open boundary conditions, usually  in time direction,  has been studied extensively as a method to overcome this difficulty. However, our preliminary study of topological charge per time slice $Q(t,\tau)$ measured by 5Li
definition after Wilson flow with flow time $t_f=32$,
as shown in Figure \ref{fig:top_t}, suggests 
the effect of the open boundary on the decorrelation the topological charge may be limited to the small number of time slices near the boundary, and the effectiveness of open boundaries in the ensemble generation for the lattice spacings in the range of $1/a \sim $ 3-5Gev is not clear.

Here we report on our attempt to use  Riemannian Manifold Hybrid Monte Carlo(RMHMC) to achieve Fourier acceleration of slow moving modes in HMC.
Originally proposed by  Duane and  Pendleton\cite{Duane}, and developed independently by Girolami and Calderhead\cite{Girolami} later, RMHMC makes it possible to use a momentum dependent `mass term', for the gauge momenta, in a gauge theory where field dependent operators are needed to identify low- and high- momentum modes. 
Section~\ref{sec:RMHMC} describe the details of the RMHMC implementation for Lattice QCD.
Analysis of HMC forces from the gauge and fermion parts of the Hamiltonian with the Laplace operator is described in Section~\ref{sec:force}. 
Section~\ref{sec: 211f} describe the result of our preliminary study of application of RMHMC on dynamical DWF ensembles.
We summarize our results in Section~\ref{sec:conclusion }.

\section{Riemannian Manifold Hybrid Monte Carlo(RMHMC)}
\label{sec:RMHMC}

In \cite{Duane}, a perturbative analysis of the Wilson gauge action showed the modes with low spacetime momentum are also slowly varying in MD time due to the smallness of the `HMC force', or the derivative of the action with respect to the gauge links. 
In principle this makes applying Fourier Acceleration techniques attractive. However, 
the gauge symmetry of QCD makes it necessary to employ a field dependent operator, the gauge covariant Laplace operator in our case, to distinguish low- and high- momentum modes by the eigenvalues of the Laplace operator.  An alternative approach using `soft' gauge fixing is also being investigated\cite{GFHMC}.

The presence of the field dependent operator in the momentum part of the Lattice QCD Lagrangian 
makes the Hamiltonian non-separable and explicit  integrators traditionally used for LQCD no longer satisfy reversibility and symplecticity, necessary for maintaining detailed balance. This can be overcome by employing implicit integrators, as noted in \cite{Girolami}.

Also, the field dependent mass term for the gauge momenta generate the determinant which has to be canceled to recover original probability distribution.
This is achieved by introducing auxiliary fields and their corresponding momenta, denoted by $\phi$ and $\pi$ respectively.
These fields are essentially duplicates of the gauge momenta, except for the mass term applied to the momenta which is the inverse of that for the gauge field. Putting it all together, the QCD Hamiltonian for RMHMC used in this study is

\begin{gather}
\HUp =  \SU
+ \frac12 \sum_\mu \left[\bp^\dagger \MU^{-1} \bp\right]
+ log \left| M \right |  \nonumber \\ 
=
{\bf S_G(U)} + {\bf S_F(U)} + \frac12 \sum_\mu \left[\bp^\dagger \MU^{-1} \bp
 + \pi_\mu^\dagger \MU \pi_\mu
+ \phi_\mu^2  \right]
\label{eq:action} 
\end{gather}

Where the mass $\MU$, is a rational function of the gauge covariant Laplace operator
\begin{gather}
\MU^{-1} = \left(G[\nabla^2]\right)^2 \\
G(x) = 
  \frac{\sum_{i=0}^n \beta_i x^i  }{\sum_{i=0}^n \alpha_i x^i}
  =G_0+\sum_{i=0}^{n/2}[a'_i x + b'_i][x^2+c'_ix + d'_i]^{-1} \\
\nabla^2 \phi_\nu(x) =  \frac{1}{16}\sum_{\mu=0}^{3}
\left[
U_\mu(x)  \phi_\nu(x+\mu) U^\dagger _\mu(x)  +
U^\dagger_\mu(x-\mu)  \phi_\nu(x-\mu) U_\mu(x-\mu) - 2\phi_\nu(x) \right] 
\end{gather}

We parameterize $\MU$ as a general rational function of Laplace operator to allow for greater flexibility. The rational form of this mass term also makes it trivial to calculate the necessary inverse mass factor for the auxiliary field.

\section{Laplace operator analysis of HMC force} 
\label{sec:force}
\begin{figure}
\includegraphics[width=0.49\linewidth]{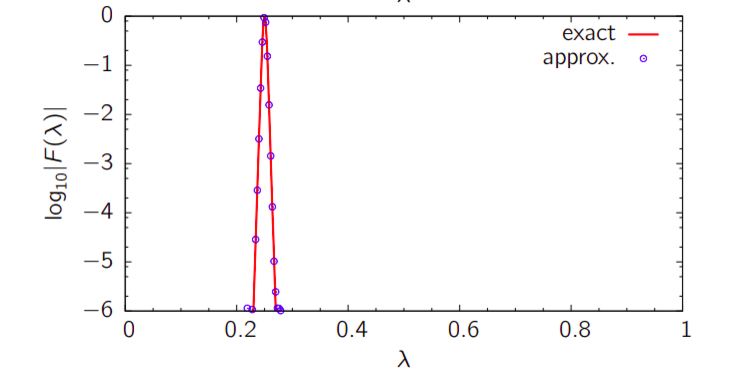}
\includegraphics[width=0.49\linewidth]{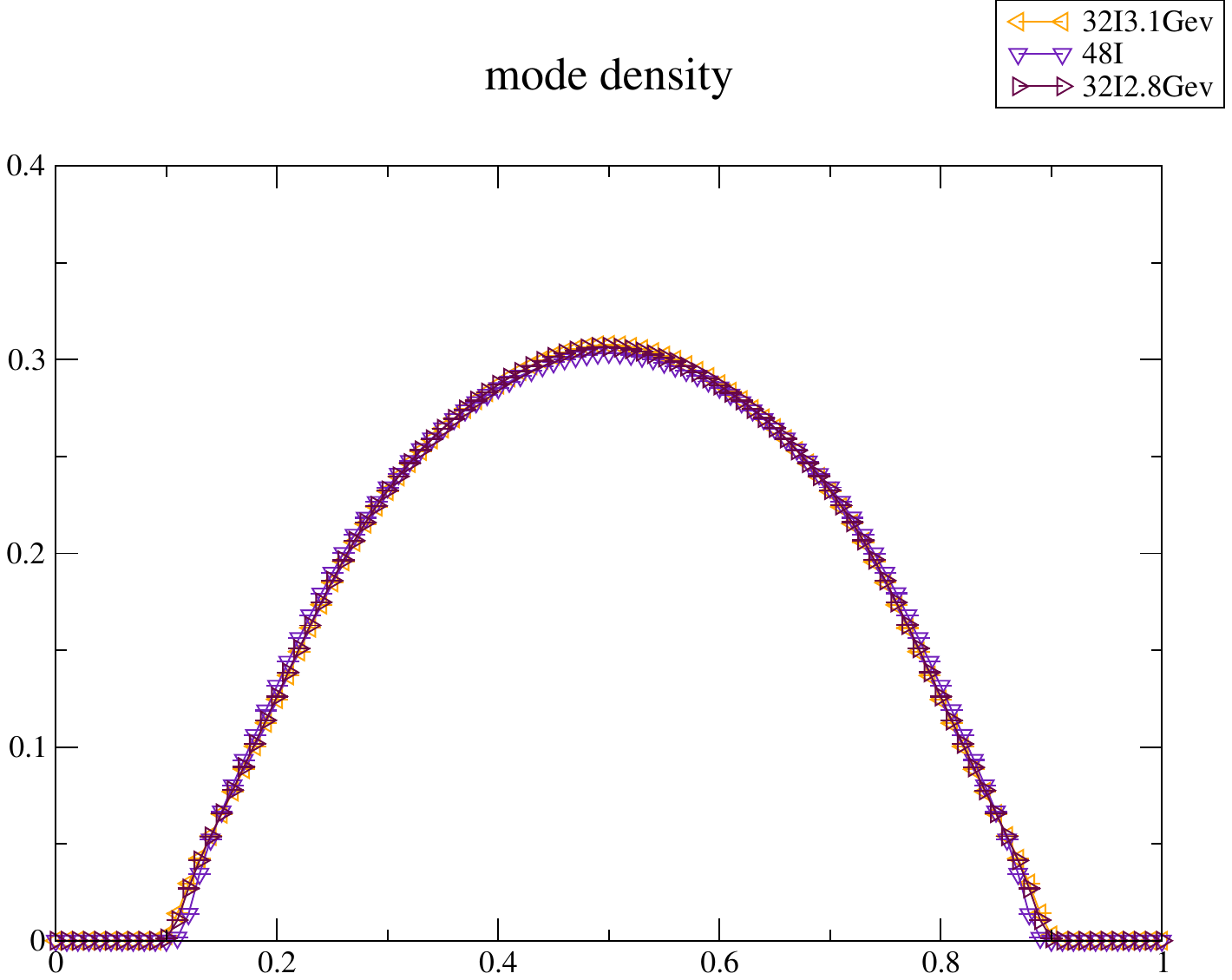}
\caption{Illustration of the bandpass filter used in the study(from Y. Jang, left) and Laplace mode density for dynamical ensembles (right).}
\label{fig:gf}
\end{figure}

Following~\cite{Tuan}, we analyze the distribution of eigenmodes of the Laplace operator, and the relative amplitude of the force from the gauge and fermion action for the eigenmodes,
lying in a narrow range, on thermalized dynamical ensembles.
For this purpose, we constructed a
Chebyshev approximation to a  sharply peaked bandpass filters centered on and eigenvalue $\lambda$, which we refer as $B_\lambda(\nabla^2(U))$.
An lllustration of the filter is given in the left panel of Figure~\ref{fig:gf}. 

First, $B_\lambda(\nabla^2(U))$ is applied to a set of gaussian random vectors to estimate mode density of 
the Laplace operator. 
The right panel of Figure~\ref{fig:gf} shows the density of modes for different ensembles.
There is a noticeable gap in the spectrum at the extreme ends for all the ensembles studied. This was true for the $\beta=10$ pure $SU(3)$ Wilson gauge ensemble with a very small lattice spacing.

\begin{figure}
	\includegraphics[width=0.49\textwidth]{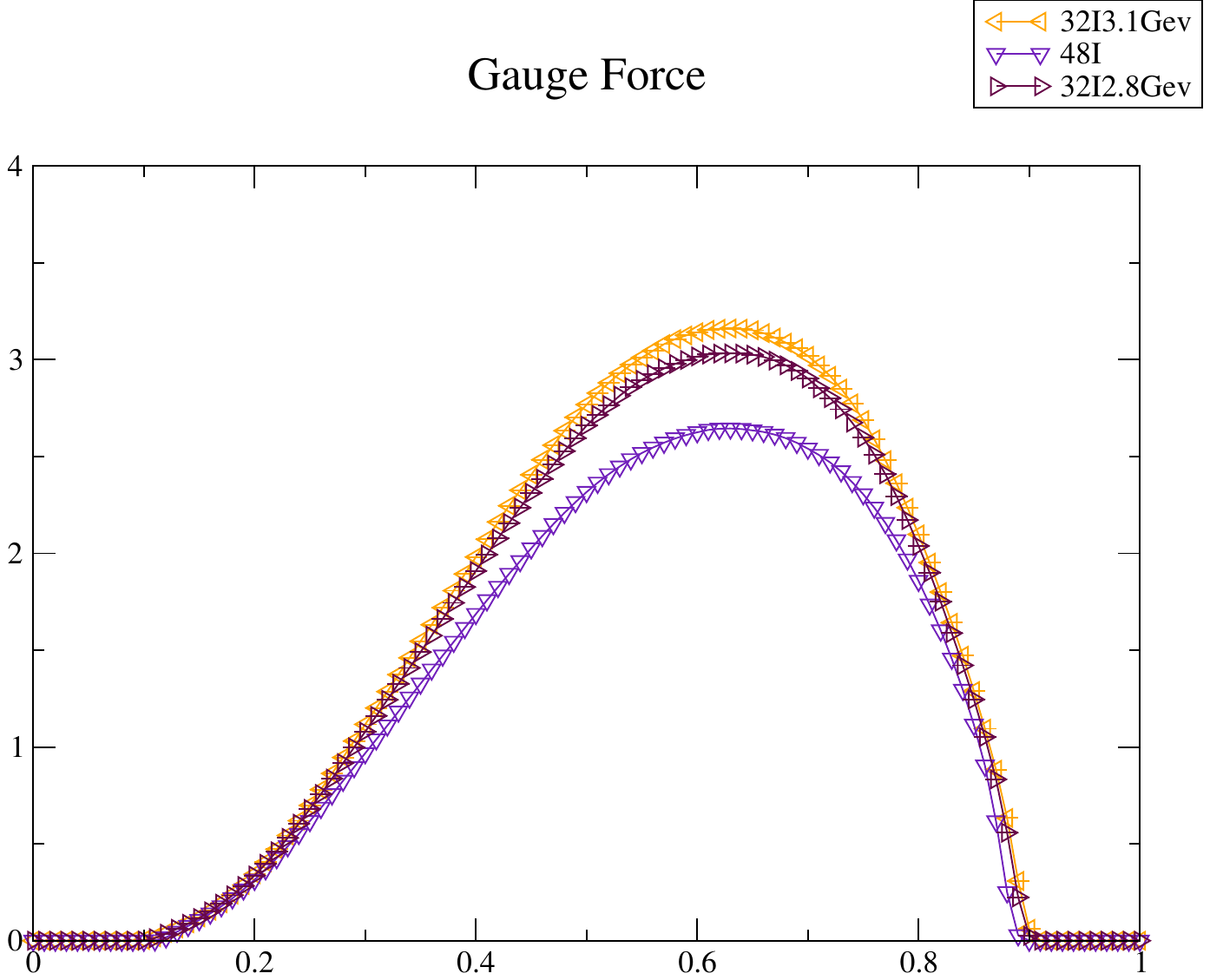}
	\includegraphics[width=0.49\textwidth]{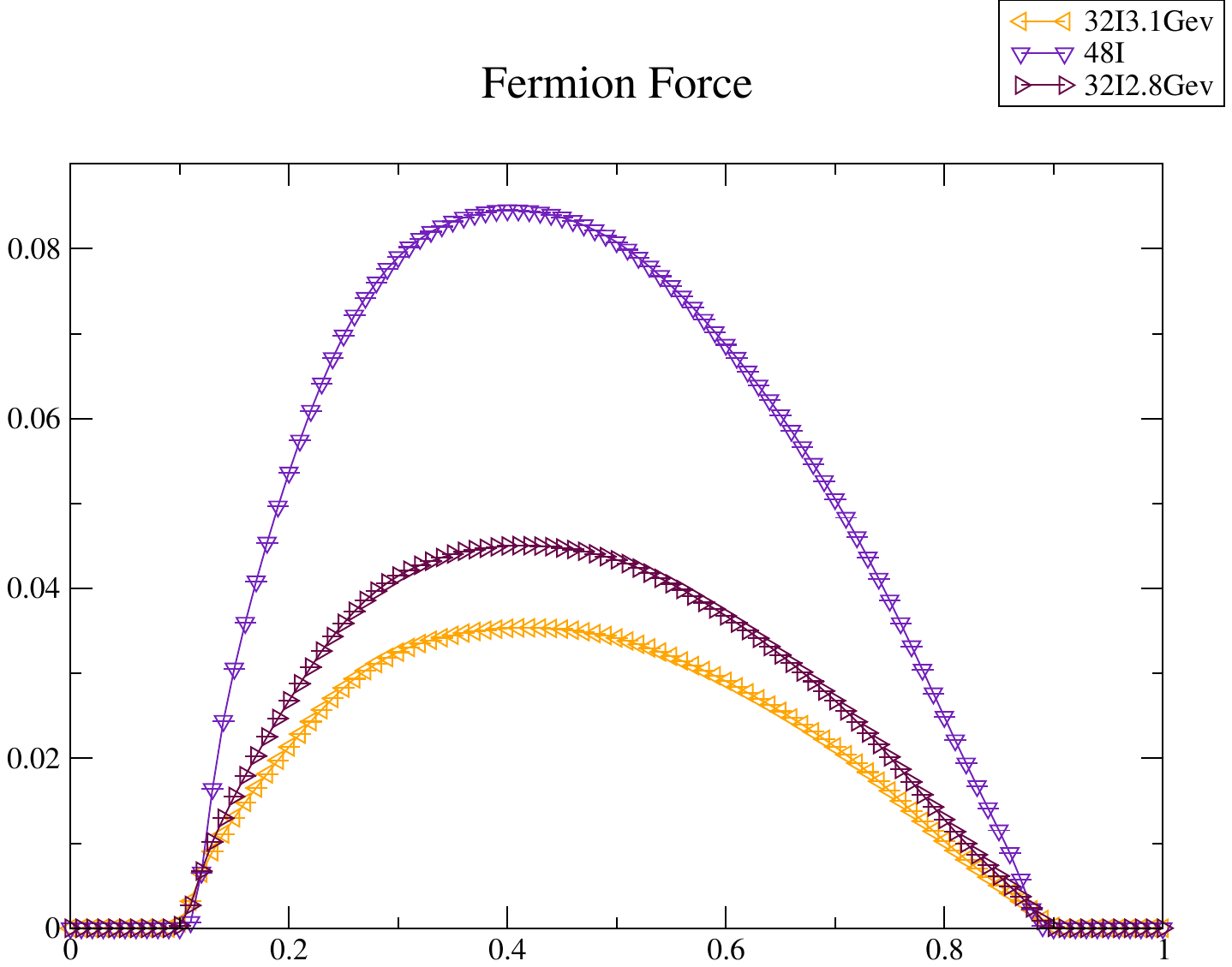}
\caption{Gauge and fermion force distributions according to the eigenvalues of  Laplace modes for 2+1-flavor DWF dynamical ensembles, at $1/a\sim$ 1.7Gev(48I), 2.8Gev, 3.1Gev respectively.}
\label{fig:ff}
\end{figure}
\begin{figure}
  \includegraphics[width=0.49\textwidth]{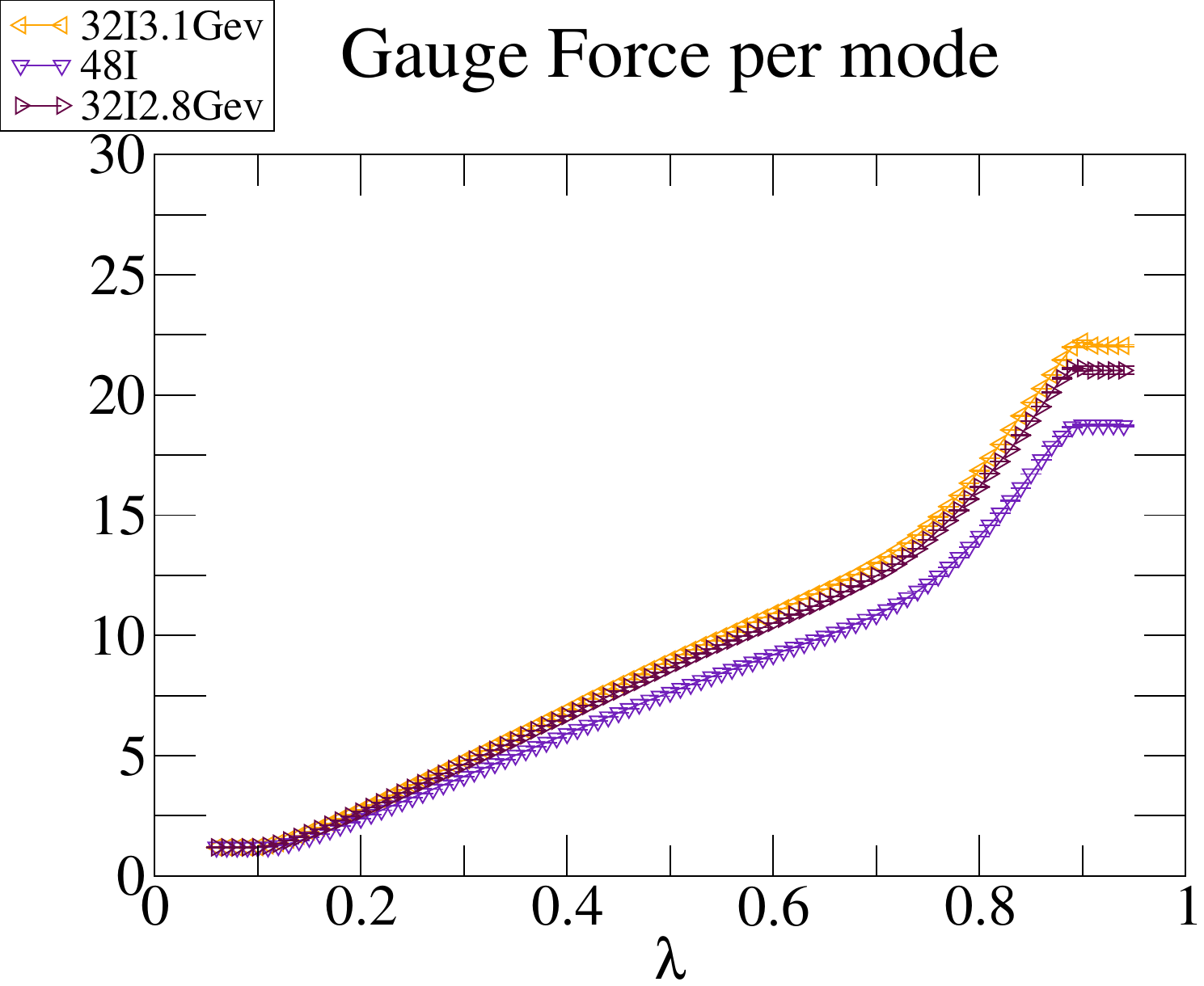}
  \includegraphics[width=0.49\textwidth]{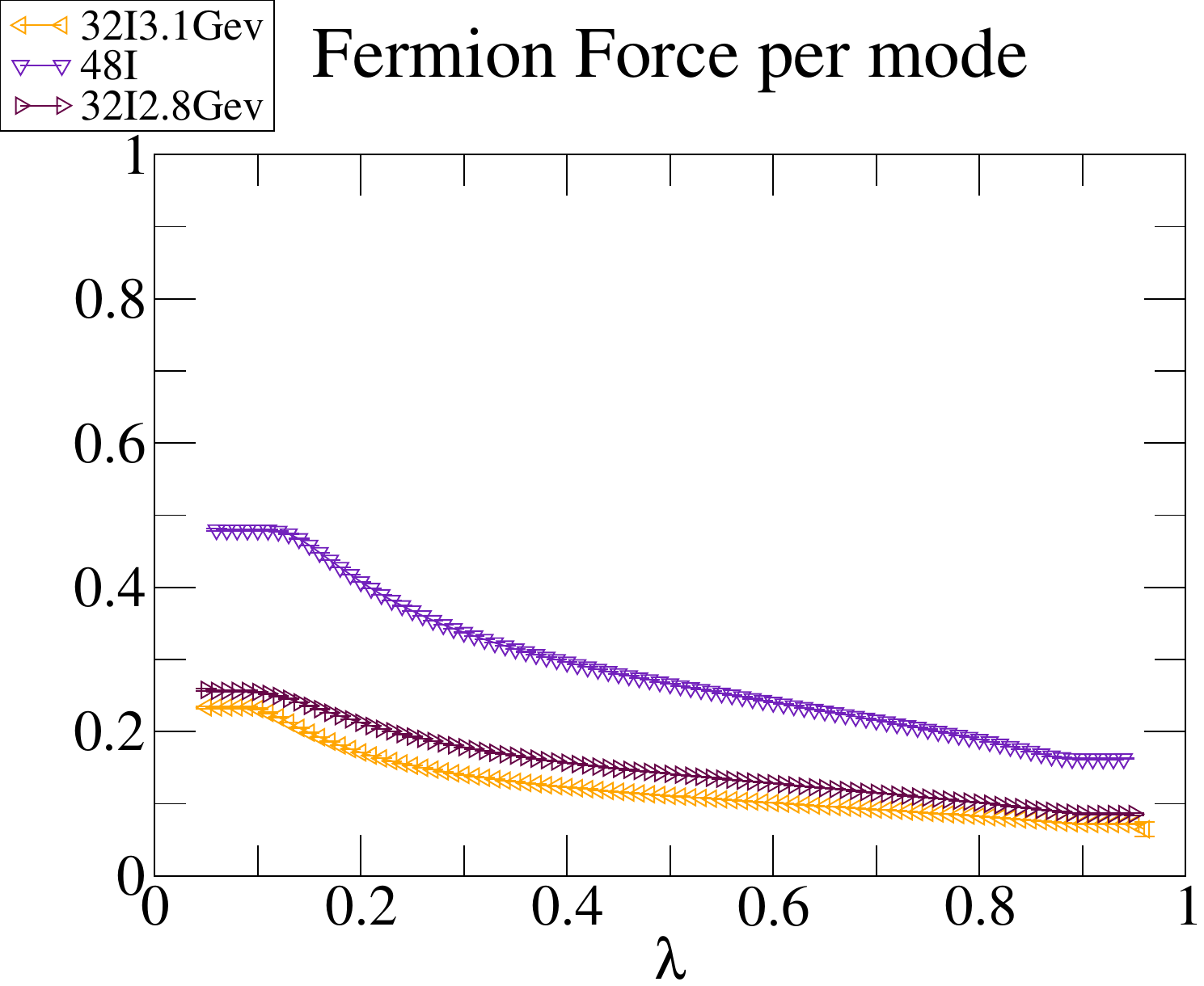}
\caption {Gauge and Fermion force density per Laplace mode for dynamical ensembles. Note the dramatic and expected difference in the eigenvalue dependence of the force generated by a local (left panel) and a non-local(right panel) action.}
\label{fig:fden}
\end{figure}

Now, the same filter is applied to the HMC forces to calculate $\left<F_{G,F}|B_\lambda(\nabla^2)|F_{G,F}\right>$ where $F_{G,F}$ is the derivative of the gauge or fermion action with resect to the gauge field $U_\mu(x)$.
Figure~\ref{fig:ff} shows the amplitudes of twith resect to he gauge and fermion forces
for different Laplace modes for various dynamical DWF ensembles. 
Now the strength of the HMC forces per Laplace mode is calculated by the ratio of the 2 quantities, as shown in Figure~\ref{fig:fden}. 

\begin{figure}
\includegraphics[width=0.52\textwidth]{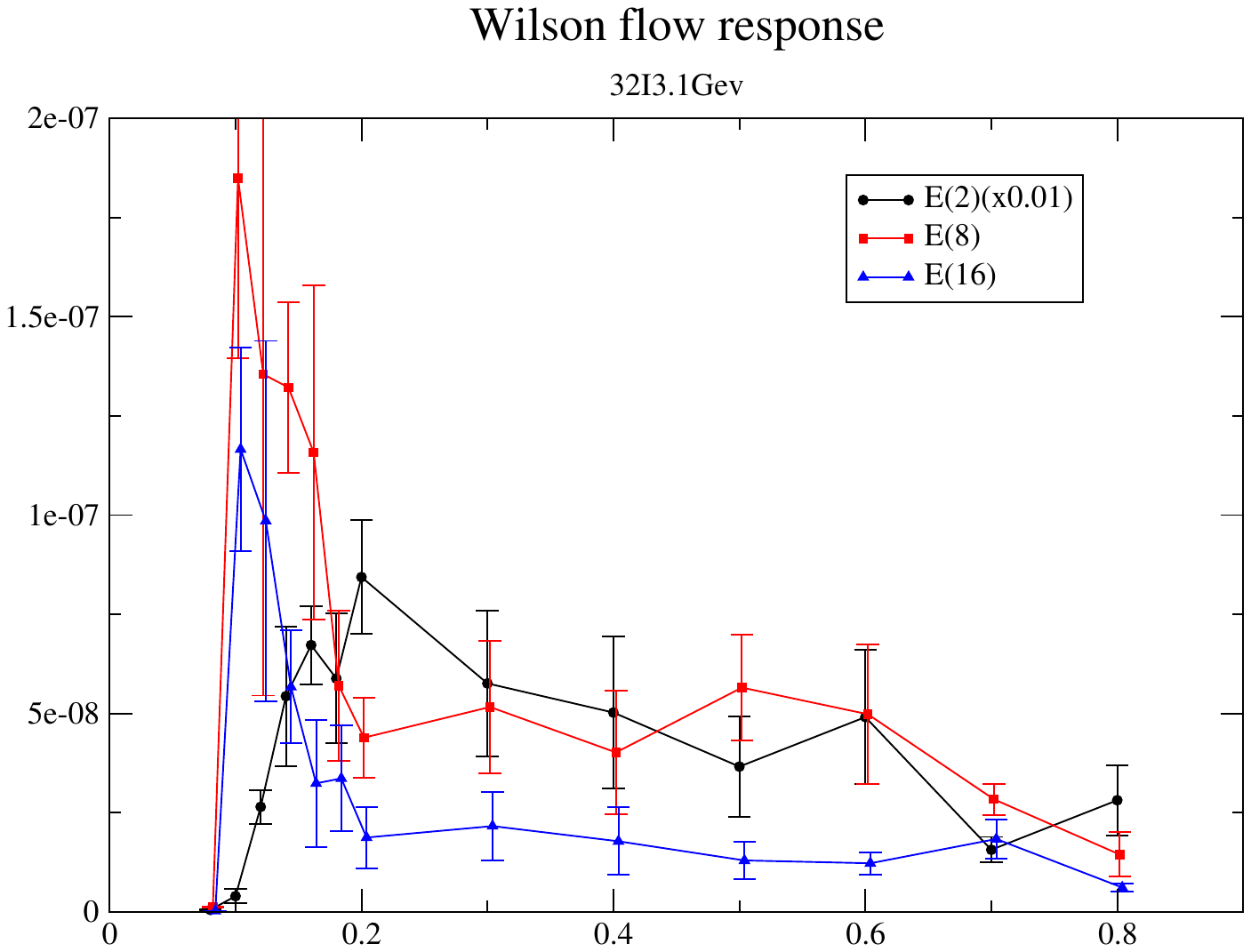}
\includegraphics[width=0.48\textwidth]{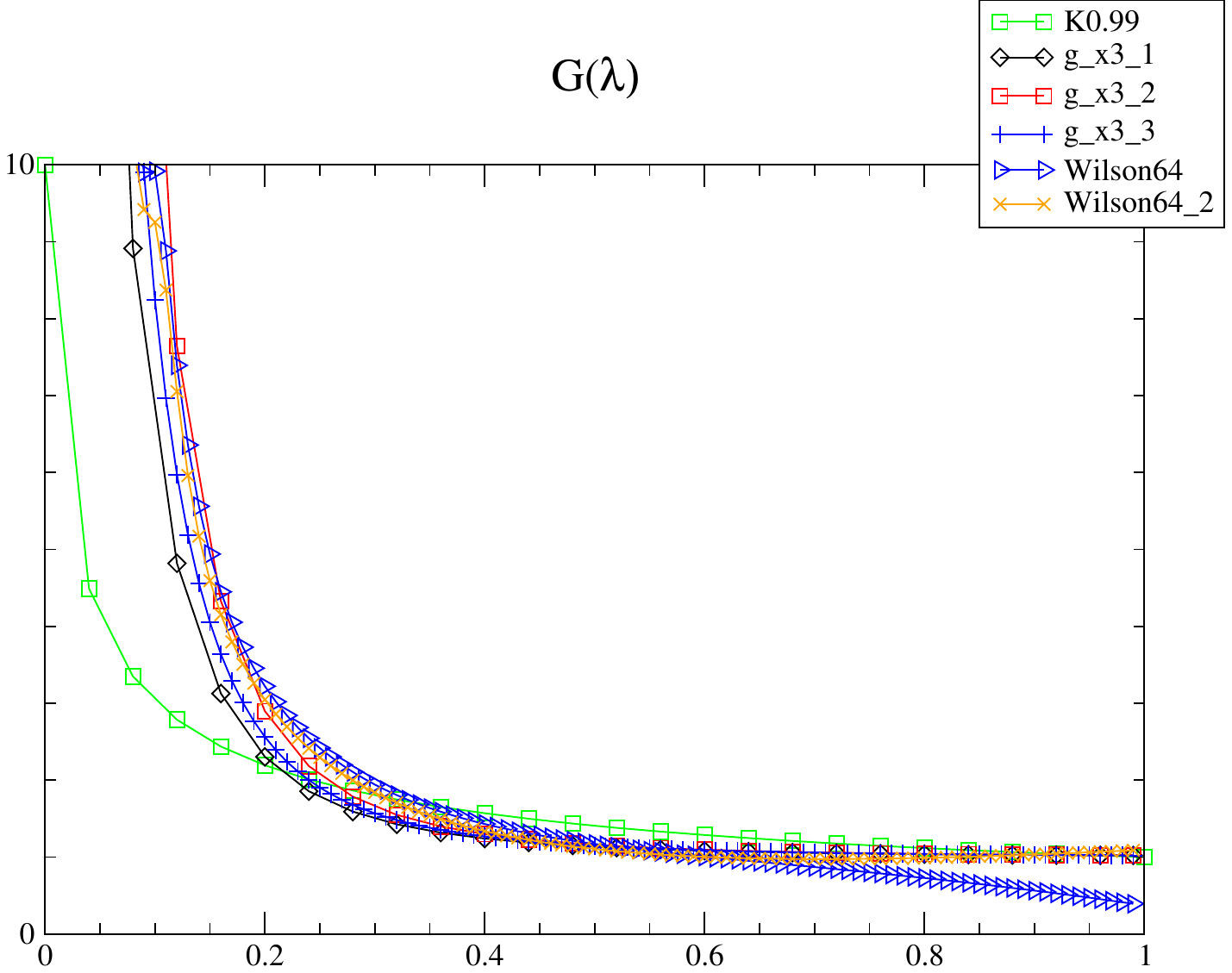}
	\caption{{\bf Left panel:} Average response of Wilson flowed energy at flow times 2,8,16 for Laplace modes on a $32^3\times 64$,  2+1-flavor ensemble with $1/a\sim 3.1$Gev(32Ifine in \cite{RBC64I}). (x0.01) in the legend denotes that the values for $E(2)$ was multiplied by that amount to be kept in the same graph as others.
	{\bf Right panel:} Numerical values of $\GU$ plitted as a function of Laplace eigenvalue. }
\label{fig:wf}
\end{figure}

Lastly, we use this filtering kernel to disentangle the effect of different Laplace modes on various observables, by the following procedure: we run a very short trajectory of HMC, with the initial momenta filtered by $B_\lambda(\nabla^2)$
centered on different eigenvalues of the Laplace operator, then measure the change in observables such as 
Wilson flowed energy $\left< E(t) \right>$ at various flow times $t$.
We found it is necessary to keep the trajectory short and use  $\tau = 10^{-4}\sim 10^{-3}$, as the initial momenta distribution, far from thermalized, rapidly diffuses toward the distribution according to the Hamiltonian. 
The left panel Figure~\ref{fig:wf} shows that the low modes of the Laplace operator correlates much more strongly with the flowed energy at larger Wilson flow time ($t\geq 8$).

Based on our findings, we have constructed mass factors of various shapes.
The right panel of Fig.~\ref{fig:wf} shows some of the mass factors $\GU$ used in this study, as well as the mass factor used in \cite{Duane}.
The mass factor Wilson64\_2 approximates the inverse of the gauge force density measured at the low momentum region if $\beta=6.4$ Wilson SU(3) ensemble, and approaches 1 in the large momentum region. The kernels we've found  to be optimal, shown as $g\_x3\_\{1,2,3\}$, have the approximate form of $1+ \frac{c'}{(x+b)^3}$.

\section{ RMHMC on dynamical DWF ensembles}
\label{sec: 211f}
\begin{figure}
\includegraphics[width=0.49\textwidth]{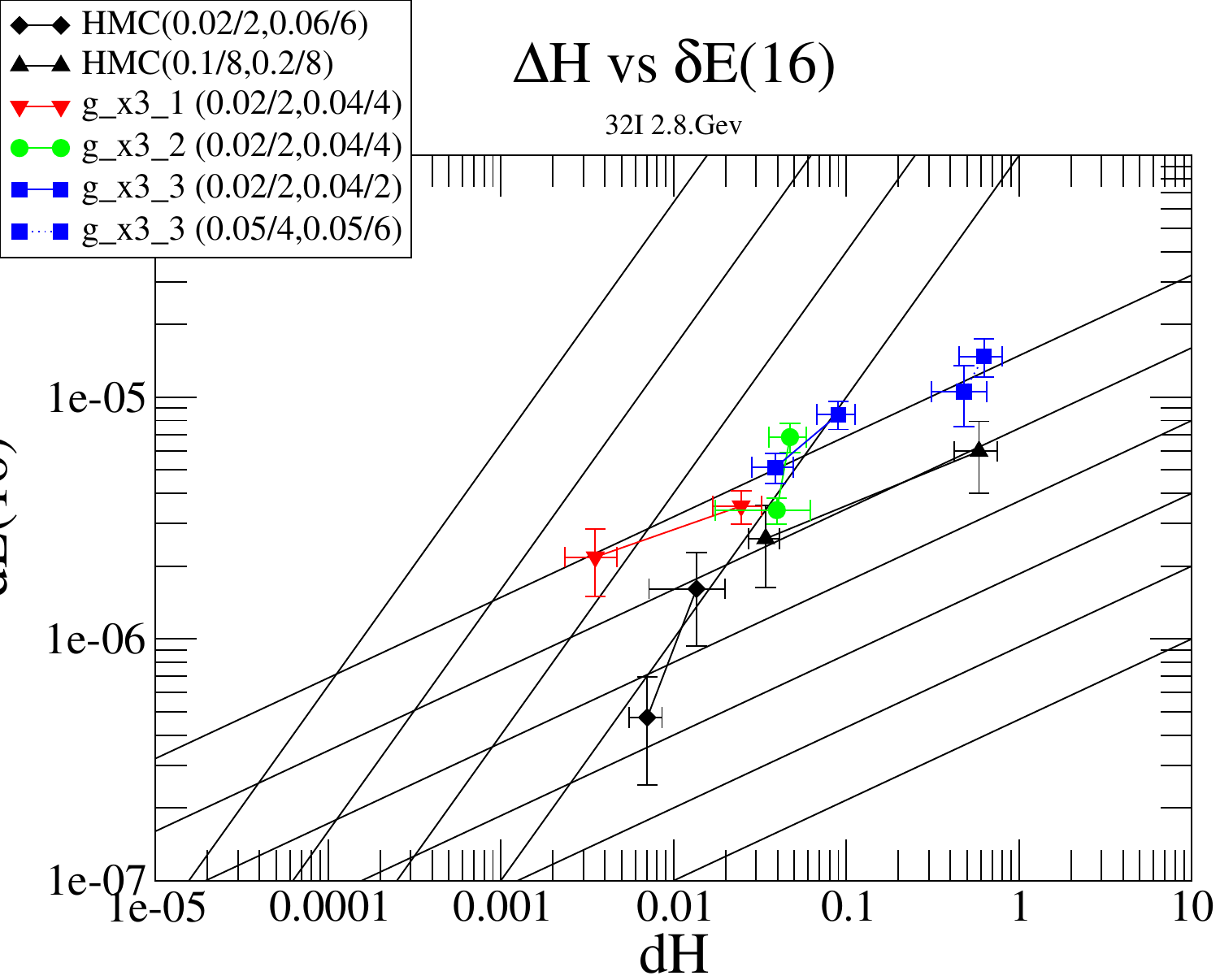}
\includegraphics[width=0.49\textwidth]{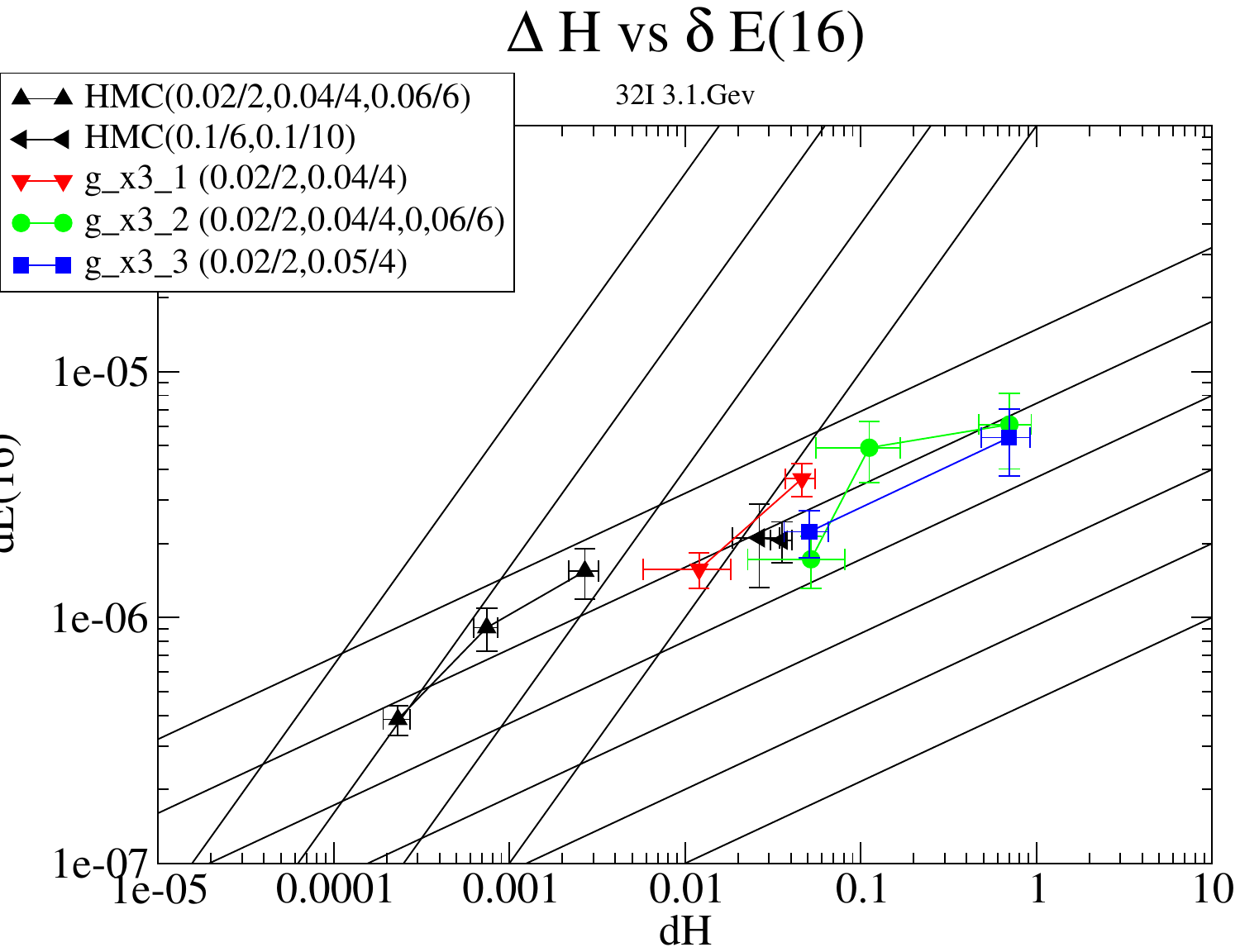}

\caption{ Average change of $E(16)$, the Wilson flowed energy at flow time 16, 
versus the average value of $\Delta H$ for a single HMC or RMHMC fermion time step on a 2+1-flavor ensembles with $1/a\sim$2.8Gev and physical quark masses(left panel) and $1/a\sim$3.1Gev . Legends denotes the mass factor, followed by (fermion step size $/$ Sexton-Weingarten factor ) for each point.  }

\label{fig:dH_dE}
\end{figure}

	In \cite{Tuan}, a significant gain in Wilson flowed energy change 
at large flow time per integrator step  was observed on a quenched ensemble with small lattice spacing (Wilson $\beta=10$). Here we repeat the exercise on realistic dynamical ensembles. Instead of gauge integrator step size, we focus on the fermion integrator step.

As a first step, we tested various mass factors on two 2+1 flavor dynamical ensembles: $1/a\sim 2.8$Gev with near physical mass, and $1/a\sim$3.1Gev, with somewhat heavier  ($m_\pi \sim$ 300Mev ) light pseudoscalar mass, with only 1 fermion step, but with different number of steps for the gauge action. 

Our tuning procedure was as follows: first, we started from very small stepsize for both gauge and fermion and checked that it generates a small enough ($\leq 10^{-2}$) $\Delta H$, to ensure other parameters such as 
the fermion inverter stopping condition 
are not affecting the tuning. After this is done, we kept the gauge step size 
fixed while increasing the fermion step size (and the overall trajectory length) by adjusting the Sexton-Weingarten ratio. 
As we increased the fermion step size, 
we observed a change in the  scaling between stepsize and $\Delta H$: when the fermion step size is small and gauge action dominates the integrator error, $\Delta H$ increase linearly, in proportion to the number of gauge steps. 
As the fermion step size increases further, integrator error from fermion action starts to dominate, and $\Delta H$ starts to scale as $\Delta t ^\delta (\delta \geq 3) $ as you would expect from single-step integration. 
The change in Wilson flowed energy is expected to be linear in the overall trajectory length.

Figure~\ref{fig:dH_dE} clearly exhibits the transition between these two scaling behaviors. The two sets of lines with different slopes represent $\Delta H \sim \delta E(16) $ and $\delta H \sim (\delta E(16))^3$, 
and adjacent parallel lines differ by a factor of 2.
The fermion step size where this change in scaling occurs may be considered an optimal choice.
Despite the relative similarity in terms of the number of flavors and lattice spacing, 
for the heavier quark ensemble(32I3.1Gev), the optimal fermion step size for inceasing $\delta E(16)$ is similar to single step HMC, although at a nominally different step size. On the other hand, for the $1/a\sim 2.8$Gev physical mass ensemble, the RMHMC appears to gain about a factor of 2 in $\delta E(16)$ per fermion step  compared to the HMC.

\begin{figure}{H}
\includegraphics[width=0.49\textwidth]{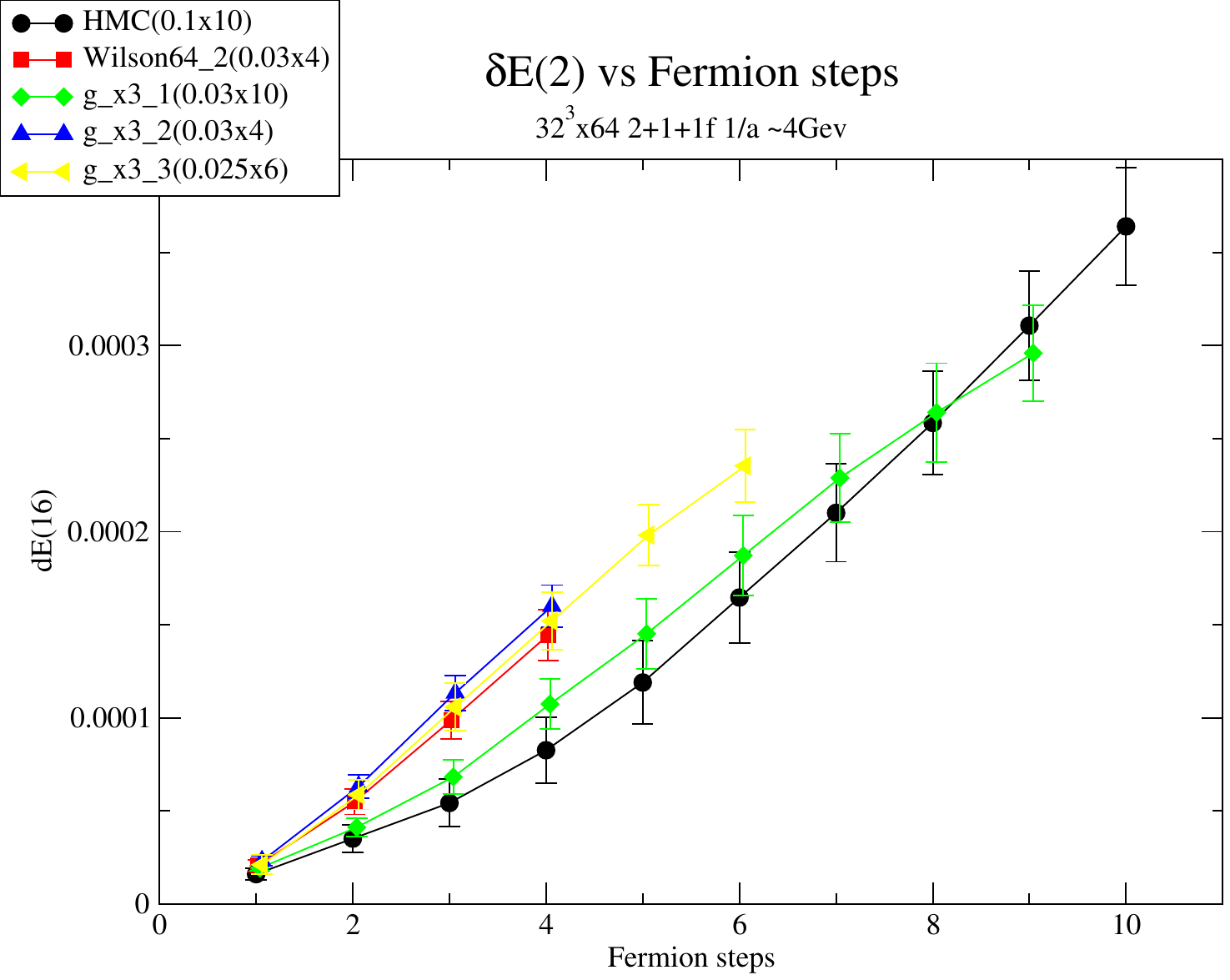}
\includegraphics[width=0.49\textwidth]{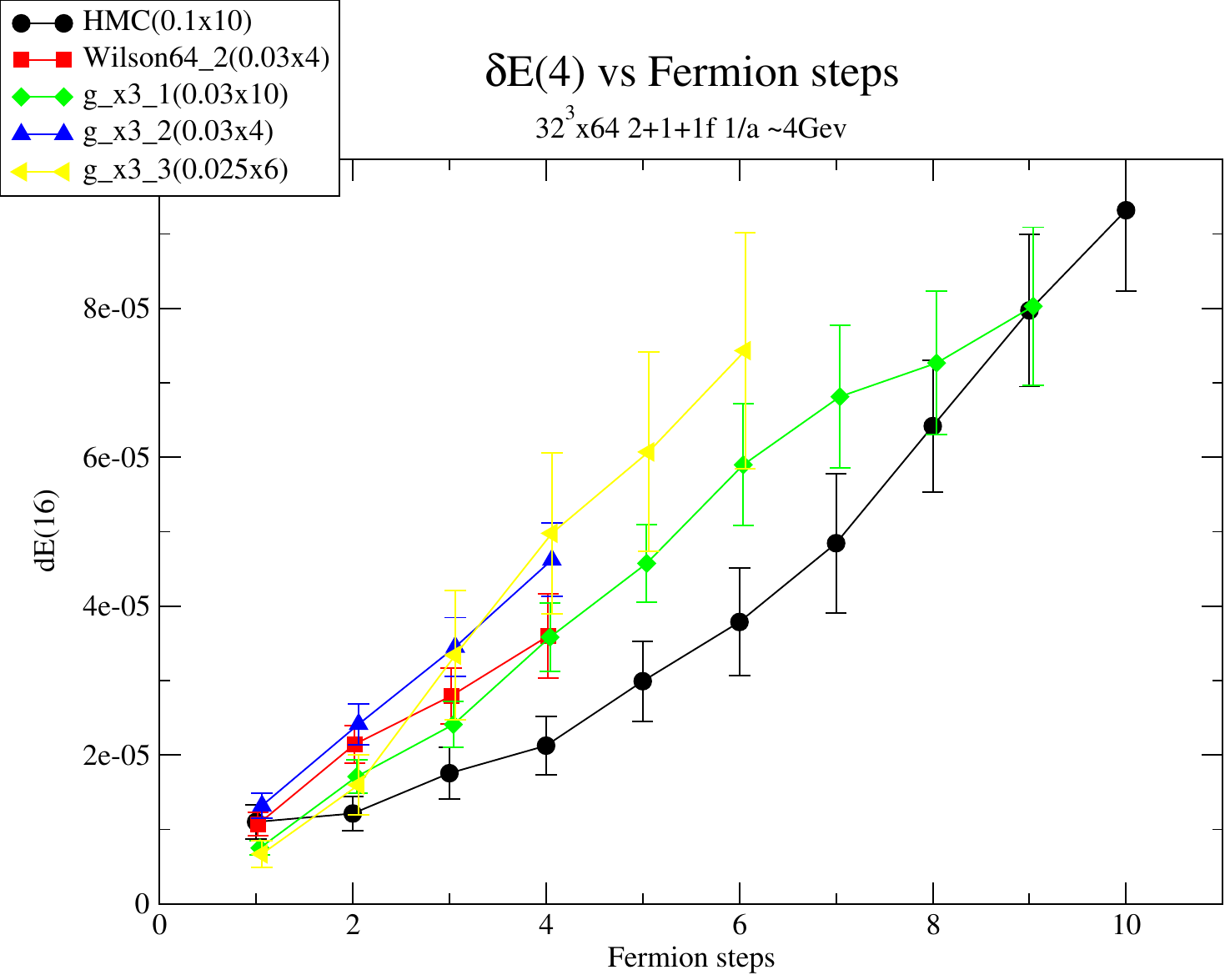}

\includegraphics[width=0.49\textwidth]{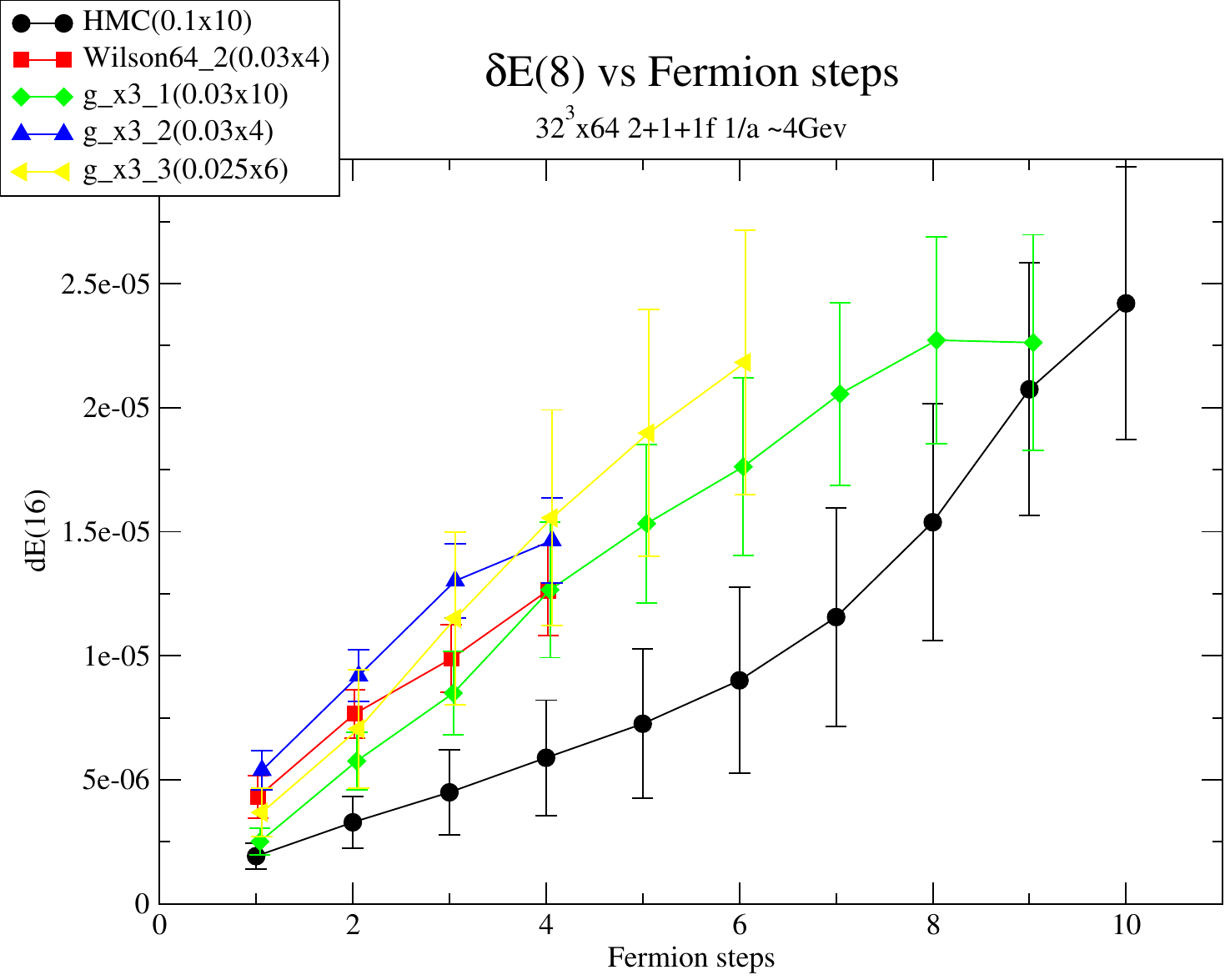}
\includegraphics[width=0.49\textwidth]{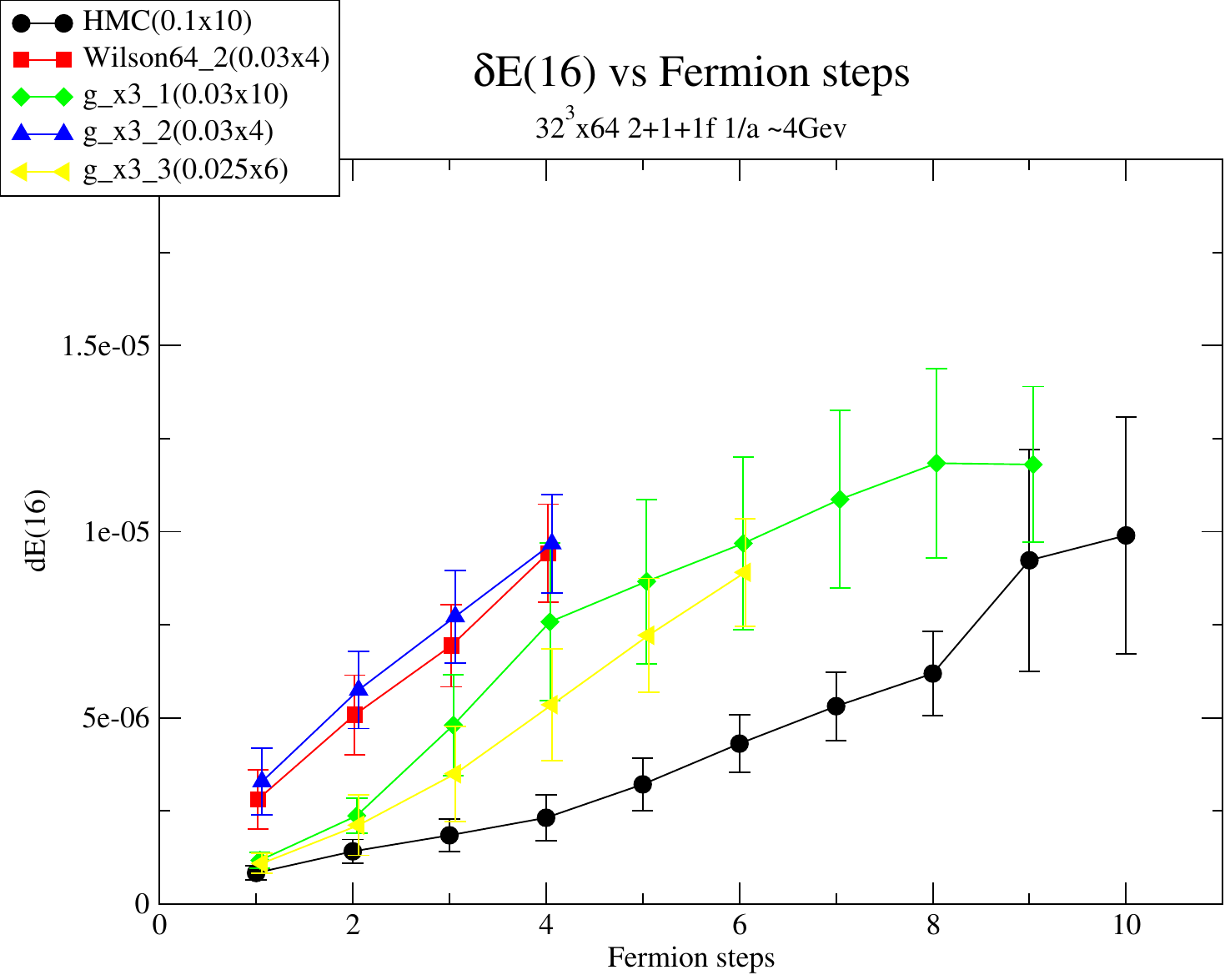}

\caption{The average change in the Wilson flowed energy at flow time 2,4,8,16 for each fermion integration step for the HMC and RMHMC with different rational functions of the Laplace operator on a 2+1+1 flavor DWF dynamical ensemble with $1/a\sim$4Gev. 
The legends denote the mass factors shown in the right panel of Fig.~\ref{fig:wf}, followed by the step size and the number of steps for the fermion integrator.} 
\label{fig:211f4Gev}
\end{figure}

Following this, we applied the same method to optimize the RMHMC  for a reduced-volume version 
of one of the ensembles we are generating as a part of the RBC/UKQCD 2+1+1 flavor DWF ensemble generation program, at $1/a\sim$4Gev. 
As Figure~\ref{fig:211f4Gev} shows, while the change in the Wilson flowed energy flowed for a shorter time does not change noticeably between the HMC and the RMHMC, $E(8)$ and $E(16)$ shows a clear gain in the RMHMC compared to the HMC.

\section{ Summary and Discussions  }
\label{sec:conclusion }

Reducing critical slowing down in the HMC is becoming a critical issue in controlling the computational cost of generating statistically independent gauge field configurations.  
We have studied the Riemannian Manifold HMC with the mass factor constructed from the gauge covariant Laplace operators, and compared the effectiveness in changing 
long distance modes identified by the Wilson flowed energy density.
It was observed that the low modes of Laplace operator strongly correlate 
with long range modes probed by 
the Wilson flowed energy, and that a hand-made mass factor achieves a factor of 2 increase in the change in Wilson flowed energy per fermion step for a 2+1+1 flavor dynamical ensemble with near physical quark mass and $1/a\sim 4$Gev.

This is one of the first cases of utilizing an implicit integrator for the QCD Hamiltonian and some observations  may be useful for future studies: While the functional form of the mass term employed in this study were reasonably well behaved in terms of Conjugate Gradient(CG) iterations needed to converge each term of the rational function, each implicit integrator step required 5-6 iterations to accurately determine the fixed point.
Also, a preliminary study of using mixed precision solvers for the rational function indicate the lower precision of inner CG interferes with the implicit integrator step convergence, and a tighter stopping condition was needed to maintain reasonable acceptance. An alternative implementation of the RMHMC which avoids fixed-point iterations necessary for the implicit integrator may prove to be beneficial, although there are theoretical issues to be understood.

We are continuing our studies of the RMHMC with different functional forms for the mass factor on 
$1/a \sim 4$Gev and finer ensembles.  While the Laplace operator is a natural choice for the operator to identify low momentum modes,
it is far from  a unique one.
The hessian of the gauge action itself is also an natural choice. The investigation of RMHMC with the hessian operator will continue as a part of US Lattice SciDAC activities.

Lastly, it should be noted that there was an error in the implementation of the implicit Omelyan integrator used in this study. We reproduce some of the runs used in this study with the corrected implementation, and the change from this error is small and does not change the qualitative conclusion drawn here.

\vspace{0.2in}

\noindent
{\bf Acknowledgment}\\
We thank our RBC and UKQCD Collaboration colleagues, Yong-Chull Jang, Peter Boyle, Bob Mawhinney in particular, for helpful discussions and ideas. 
This research was supported by the Exascale
Computing Project (17-SC-20-SC), a collaborative effort of the U.S. Department of Energy Office
of Science and the National Nuclear Security Administration. We are grateful for the computational
resources provided by Argonne Leadership Computing Facility, which is a DOE Office of Science
User Facility supported under Contract DE-AC02-06CH11357 and the Oak Ridge Leadership
Computing Facility at the Oak Ridge National Laboratory, which is supported by the Office of
Science of the U.S. Department of Energy under Contract No. DE-AC05-00OR22725. In addition
NHC was supported by U.S. DOE grant No. DE-SC0011941. RMHMC implementation used in this study are based upon
Grid(https://github.com/paboyle/Grid) and CPS(https://github.com/RBC-UKQCD/CPS\_public).

\end{document}